\documentclass[dvips,12pt]{article}
\usepackage{graphicx,color}
\newcommand\figcaption{\def\@captype{figure}\caption}
\newcommand\tabcaption{\def\@captype{table}\caption}

\title{\Large \bf A D-dimensional Heckmann-like solution of Jordan-Brans-Dicke
theory}
\author{S.M.Kozyrev\footnote {email: Sergey@tnpko.ru}}
\date{\small Scientific center gravity wave studies "Dulkyn", Kazan, Russia}

\begin{document}
\maketitle\DeclareGraphicsRule{.eps.gz}{eps}{.eps.bb}{`gunzip -c
#1}

\begin{abstract}
In this short letter we present a some rigorous vacuum solutions
of the D-dimensional Jordan-Brans-Dicke field equations. In
contrast with the well known Brans-Dicke solutions \cite{Brans}, to the search
of static and spherically symmetric space-time we choose the
widespread Hilbert coordinates.


\end{abstract}

The continuing focus on extra-dimensional models in fundamental
physics provides motivation for studying scalar fields in higher
dimensions.The purpose of this short letter is to demonstrate that
one can find the solutions in widespread Hilbert coordinates by
exploiting the change of variables technique. First time, this
technique was applied to a spherically symmetric case for finding
a new solution to the Jordan-Brans-Dicke scalar field equations
\cite{Jor1}. Techniques for obtaining the similar static solutions
are known by know \cite{Koz1,Koz2}. In D = d + 1 dimensions, the
field equations given by (we take units G = c = 1):
\begin{eqnarray}
R_{\mu \nu }-\frac 12Rg_{\mu \nu }=\frac \omega {\phi ^2}\left(
\phi _{,\mu }\phi _{,\nu }-\frac 12g_{\mu \nu }g^{\alpha \beta
}\phi _{,\alpha }\phi _{,\beta }\right) +\frac 1\phi \left( \phi
_{\mu \nu }-g_{\mu \nu }\phi \right) ,
\end{eqnarray}

\begin{eqnarray}
\nabla ^\alpha \nabla _\alpha \phi =0.
\end{eqnarray}
$\nabla _\alpha $ is the covariant derivative associated with the
metric g, and $\nabla ^\alpha \nabla _\alpha $ is the D
dimensional Laplace operator of this metric, R$_{\mu \mu }$ and R
are the Ricci tensor and Ricci scalar for an arbitrary metric g.
As we have already mentioned we consider standard static and
spherically symmetric space- time in Hilbert coordinates for d + 1
dimensions:
\begin{eqnarray}
ds^2=-e^{\nu (r)}dt^2+e^{\lambda (r)}dr^2+r^2d\Omega _d^2,
\end{eqnarray}

where $\lambda $, $\nu $ are unknown functions of the radial
coordinate r, and d$\Omega _d^2$ is the solid angle element in d -
1 dimensions. In order to simplify the problem of solving the
field equations we will replace variable r by r($\nu $) then the
field equations take a form:
\begin{eqnarray}
1+2\left( d-1\right) \frac{r^{\prime }\left( \nu \right) }{r\left(
\nu \right) }-\lambda ^{\prime }\left( \nu \right)
-\frac{2r^{"}\left( \nu \right) }{r^{\prime }\left( \nu \right)
}+\frac{2\phi ^{\prime }\left( \nu \right) }{\phi \left( \nu
\right) }=0,
\end{eqnarray}
\begin{eqnarray}
-1+\lambda ^{\prime }\left( \nu \right) \left( 1+2\left( d-1\right) \frac{%
r^{\prime }\left( \nu \right) }{r\left( \nu \right) }+\frac{2\phi
^{\prime }\left( \nu \right) }{\phi \left( \nu \right) }\right)
+\left( 2+\frac{4\phi
^{\prime }\left( \nu \right) }{\phi \left( \nu \right) }\right) \frac{%
r^{"}\left( \nu \right) }{r^{\prime }\left( \nu \right) }-4\left( \frac{%
\omega \phi ^{\prime }\left( \nu \right) ^2}{\phi \left( \nu \right) ^2}+%
\frac{\phi "\left( \nu \right) }{\phi \left( \nu \right) }\right)
=0,
\end{eqnarray}
\begin{eqnarray}
-1+2\left( d-2\right) \left( e^{\lambda \left( \nu \right) }-1\right) \frac{%
r^{\prime }\left( \nu \right) }{r\left( \nu \right) }+\lambda
^{\prime }\left( \nu \right) -\frac{2\phi ^{\prime }\left( \nu
\right) }{\phi \left( \nu \right) }=0,
\end{eqnarray}
and the equation of motion for the scalar field
\begin{eqnarray}
1+2\left( d-1\right) \frac{r^{\prime }\left( \nu \right) }{r\left(
\nu \right) }-\lambda ^{\prime }\left( \nu \right)
-\frac{2r^{"}\left( \nu \right) }{r^{\prime }\left( \nu \right)
}+\frac{2\phi ^{"}\left( \nu \right) }{\phi ^{\prime }\left( \nu
\right) }=0,
\end{eqnarray}
where $\nu $ is a new variable and the primes denote derivatives
with respect to $\nu $. Making use of equation (6) in (7) they
simplify to
\begin{eqnarray}
\frac{\phi ^{"}\left( \nu \right) }{\phi ^{\prime }\left( \nu \right) }-%
\frac{\phi ^{\prime }\left( \nu \right) }{\phi \left( \nu \right)
}=0,
\end{eqnarray}
thus we obtain
\begin{eqnarray}
\phi \left( \nu \right) =\alpha e^\beta ,
\end{eqnarray}
where $\alpha $ and $\beta $ is a arbitrary constants. Using the
asymptotic
condition in infinity we have $\phi \left( \nu \right) $ =1. In the case $%
\beta \ $ = 0 one can find solution of equations (4) - (7)
\begin{eqnarray}
r\left( \nu \right) =\frac{const}{e^\nu -1},\ e^{\lambda \left(
\nu \right) }=e^{-\nu },\ \phi \left( \nu \right) =1,
\end{eqnarray}

that identical to the Schwarzschild solution of the Einstein
theory. For the more general case $\beta \neq 0$ making use
equations (4) , (6) and (9) we eliminate $\lambda ^{\prime }\left(
\nu \right) $ and obtain for r $\left( \nu \right) $

\begin{eqnarray}
r\left( \nu \right)  &=&\xi e^{-\frac{2\beta \left( \omega \beta -1\right) }{%
\Phi ^2+\Psi ^2}\left( \Phi \zeta +\left( 1+2\beta \right) \nu
+2\arctan
h\left[ \frac{\Psi \tan \left[ \frac 12\left( \zeta +\frac \nu {\sqrt{d-1}%
}\right) \right] }\Phi \right] \right) }\  \\
&&\ \left( \Phi ^2-\Psi ^2+\left( \Phi ^2+\Psi ^2\right) \right)
\cos \left[ \left( \zeta +\frac \nu {\sqrt{d-1}}\right) \Psi
\right] ^{\frac{2\beta \left( \omega \beta -1\right) }{\Phi
^2+\Psi ^2}},
\end{eqnarray}
where $\zeta $ and $\xi $ is a arbitrary constants, and
\begin{eqnarray}
\Phi  &=&\left( 1+2\beta \right) \sqrt{d-1},\  \\
\Psi  &=&\sqrt{1+4\beta \left( 2\beta \omega +\beta -1\right)
-d\left( 1+4\beta ^2\left( 1+\omega \right) \right) },
\end{eqnarray}

After same algebra one can find from (5) - (6) for $\lambda \left(
\nu \right) :$

\begin{eqnarray}
\lambda \left( \nu \right) =\ln \left[ -\frac{\Psi ^2\sec \left( \frac{%
\left( \sqrt{d-1}\zeta +\nu \right) \Psi }{2\sqrt{d-1}}\right)
^2}{4\left( d-2\right) \left( \beta \omega -1\right) }\right] ,
\end{eqnarray}

Notice that for some special circumstances one can express these
solutions in more natural view concerning a variable r. First let
us consider that we can express the field equations for the case
$\phi =\alpha e^{\frac \nu \omega },$ then

\begin{equation}
\begin{array}{l}
e^\nu =\chi \ r^{-\frac{d\omega }{2+\omega }}\left( \left(
2-d\right) r^d\sigma \omega +\left( 2+\omega \right) r^2\right)
^{\frac \omega
{2+\omega }}, \\
\\
e^\lambda =\frac 1{1-\frac{2+\omega }{\left( d-2\right) \sigma \omega }%
r^{2-d}}, \\
\\
\phi =\alpha e^{\frac \nu \omega }.
\end{array}
\label{e19}
\end{equation}

or

\begin{equation}
\begin{array}{l}
e^\nu =\left( \chi \ +\left( \frac r\sigma \right) ^{2-d}\right)
^{\frac
\omega {2+\omega }}, \\
\\
e^\lambda =\frac \chi {\chi -\left( \frac r\sigma \right) ^{2-d}}, \\
\\
\phi =\alpha e^{\frac \nu \omega }.
\end{array}
\label{e19}
\end{equation}

where $\chi $ and $\sigma $ is a arbitrary constants. On the other
hand one can write the $\phi =\alpha e^{-\frac \nu 2},$ then

\begin{equation}
\begin{array}{l}
e^\nu =const\left( \left( \frac r\sigma \right)
^{2-d}+\sqrt{1+\left( \frac r\sigma \right) ^{2\left( 2-d\right)
}}\right) ^{2\sqrt{\frac{d-1}{\left(
2+\omega \right) \left( d-2\right) }}}, \\
\\
e^\lambda =\frac 1{1+\left( \frac r\sigma \right) ^{2\left(
2-d\right) }},
\\
\\
\phi =\alpha e^{-\frac \nu 2}.
\end{array}
\label{e19}
\end{equation}

Another choice of function $\nu =a\left( b+e^{-\lambda /2}\right)
$  would lead to a different solution

\begin{equation}
\begin{array}{l}
e^\nu =e^{a\left( 1+b+\Pr oductLog\left( \eta \ r^{2-d}\right) \right) }, \\
\\
e^\lambda =\frac 1{\left( 1+\Pr oductLog\left( \eta \
r^{2-d}\right) \right)
^2}, \\
\\
\phi =\gamma e^{-\frac 12\left( a-2\right) \Pr oductLog\left( \eta
\ r^{2-d}\right) }.
\end{array}
\label{e19}
\end{equation}

where $\eta $ and $\gamma $ is a arbitrary constants and

\begin{eqnarray}
a=\frac 2{\left( 2+\omega \right) }\left( 1+\omega
+\sqrt{\frac{\omega -d\left( 1+\omega \right) }{d-2}}\right) .
\end{eqnarray}
\section*{Conclusion}

In this paper D - dimensional spherically symmetric static
solutions of the Jordan-Brans-Dicke field equations are
investigated. This analysis shows that the solutions can be
explicitly written down in Hilbert coordinates, contrary to what
it is usually claimed in the literature. A number of higher
dimensional generalizations of the Heckmann solution have been
found.

\end{document}